# Linear-Array Photoacoustic Imaging Using Minimum Variance-Based Delay Multiply and Sum Adaptive Beamforming Algorithm

**Moein Mozaffarzadeh**[a], **Ali Mahloojifar\***[a], **Mahdi Orooji**[a], **Karl Kratkiewicz**[b], **Saba Adabi**[b], **Mohammadreza Nasiriavanaki**[b]

[a]Department of Biomedical Engineering, Tarbiat Modares University, Tehran, Iran
[b]Department of Biomedical, Wayne State University, 818 W. Hancock, Detroit, Michigan, USA

**Abstract.** In Photoacoustic imaging (PA), Delay-and-Sum (DAS) beamformer is a common beamforming algorithm having a simple implementation. However, it results in a poor resolution and high sidelobes. To address these challenges, a new algorithm namely Delay-Multiply-and-Sum (DMAS) was introduced having lower sidelobes compared to DAS. To improve the resolution of DMAS, a novel beamformer is introduced using Minimum Variance (MV) adaptive beamforming combined with DMAS, so-called Minimum Variance-Based DMAS (MVB-DMAS). It is shown that expanding the DMAS equation results in multiple terms representing a DAS algebra. It is proposed to use the MV adaptive beamformer instead of the existing DAS. MVB-DMAS is evaluated numerically and experimentally. In particular, at the depth of 45 $mm$ MVB-DMAS results in about 31 $dB$, 18 $dB$ and 8 $dB$ sidelobes reduction compared to DAS, MV and DMAS, respectively. The quantitative results of the simulations show that MVB-DMAS leads to improvement in full-width-half-maximum about 96 %, 94 % and 45 % and signal-to-noise ratio about 89 %, 15 % and 35 % compared to DAS, DMAS, MV, respectively. In particular, at the depth of 33 $mm$ of the experimental images, MVB-DMAS results in about 20 $dB$ sidelobes reduction in comparison with other beamformers.

\*Ali Mahloojifar, mahlooji@modares.ac.ir

## 1 Introduction

Photoacoustic imaging (PAI) is a promising medical imaging modality that uses a short electromagnetic pulse to generate Ultrasound (US) waves based on the thermoelastic effect.[1] Having the merits of the US imaging spatial resolution and the optical imaging contrast in one imaging modality is the main motivation of using PAI.[2] Unlike the X-ray which uses an ionizing radiation, PAI uses nonionizing waves, i.e. short laser or radio frequency (RF) pulses. In comparison with other imaging modalities, PAI has multiple advantages leading to many investigations.[3,4] PAI is a multiscale imaging modality that has been used in different cases of study such as tumor detection,[5,6] cancer detection and staging,[7] ocular imaging,[8] monitoring oxygenation in blood vessels[9] and functional imaging.[10,11] There are two techniques of PAI: Photoacoustic Tomography (PAT) and Photoacoustic Microscopy (PAM).[12,13] In 2002, for the first time, PAT was successfully used as *in vivo* functional and structural brain imaging modality in small animals.[14] In PAT, an array of elements may be formed in linear, arc or circular shape, and mathematical reconstruction algorithms are used to obtain the optical absorption distribution map of a tissue.[15–17] Most of the reconstruction algorithms are defined under an ideal imaging condition and full-view array of elements. Also, the noise of measurement system is not considered as a parameter in the reconstruction procedure. Thus, Photoacoustic (PA) reconstructed images contain inherent artifacts caused by imperfect reconstruction algorithms. Reducing these artifacts has become a crucial challenge in PA image reconstruction for different number of transducers and different properties of imaging media.[18,19] Since there is a high similarity between US and PA detected signals, many of beamforming algorithms used in US imaging can be used in PAI. Moreover, integrating these two imaging modalities has been a challenge.[20,21] Common US beamforming algorithms such as Delay-And-Sum (DAS) and Minimum variance (MV) can be used in PA beamforming with modifications.[22] These modifications in algorithms have led to use different hardware to implement an integrated US-PA imaging device. There are many studies focused on developing one beamforming technique for US and PA image formation in order to reduce the cost of imaging system.[23,24] Although DAS is the most common beamforming method in linear array imaging, it is a blind beamformer. Consequently, DAS causes a wide mainlobe and high level of sidelobes.[25] Adaptive beamformers are commonly employed in Radar and have the ability of weighting the aperture based on the characteristics of detected signals. Apart from that, these beamformers form a high quality image with a wide range of off-axis signals rejection. MV can be considered as one of the commonly used adaptive methods in medical imaging.[26–28] Over time, vast variety of modifications have been investigated on MV such as complexity reduction,[29,30] shadowing suppression,[31] using of the eigenstructure,[32,33] and combination of MV and Multi-line transmission (MLT) technique.[34] Matrone *et al.* proposed, in,[35] a new beamforming algorithm namely Delay-Multiply-and-Sum (DMAS) as a beamforming technique, used in medical US imaging. This algorithm, introduced by Lim *et al.* , was initially used in confocal microwave imaging for breast cancer detection.[36] In addition, DMAS was used in Synthetic Aperture imaging.[37] Double stage (DS-DMAS), in which two stages of DMAS is used in order to achieve a higher contrast and resolution compared to DMAS, was proposed for linear-array US and PAI.[38–40] In addition, a modified version of coherence factor (MCF) and a high resolution CF were used to provide a higher contrast and resolution in linear-array PAI, respectively, compared to conventional CF.[41,42]

In this paper, a novel beamforming algorithm, namely Minimum Variance-Based DMAS (MVB-DMAS), is introduced. The expansion of DMAS algorithm is used, and it is shown that in each term of the expansion, there is a DAS algebra. Since DAS algorithm is a non-adaptive beamformer and leads to low resolution images, we proposed to use MV instead of the existing DAS in DMAS algebra expansion. It is shown that using MVB-DMAS results in resolution improvement and sidelobe levels reduction at the expense of higher computational burden. A preliminary version of this work and its eigenspace version have been reported before.[43–45] However, in this paper, we are going to present a highly more complete description of this approach and evaluate, numerically and experimentally, its performance and the effects of its parameters.

The rest of the paper is organized as follows. In section 2, the DMAS and MV beamforming algorithms are presented. In section 3, the proposed method and the necessary modifications are explained. The numerical and experimental results are presented in section 4 and 5, respectively. The advantages and disadvantages of proposed method are discussed in section 6, and finally con-clusion is presented in section 7.

## 2 Background

### 2.1 Beamforming

When PA signals are detected by a linear array of US transducer, beamforming algorithms such as DAS can be used to reconstruct the image using the following equation:

$$y_{DAS}(k) = \sum_{i=1}^{M} x_i(k - \Delta_i),$$

(1)

where $y_{DAS}(k)$ is the output of the beamformer, $k$ is the time index, $M$ is the number of elements of array, and $x_i(k)$ and $\Delta_i$ are detected signals and corresponding time delay for detector $i$, respectively. DAS is a simple algorithm and can be used for realtime PA and US imaging. However, in linear array transducer only a few numbers of detection angles are available. In other words, a low quality image is formed due to the limited available angles in linear array transducers. DMAS was introduced in[35] to improve the image quality. DMAS calculates corresponding sample for each element of the array, the same as DAS, but before summation, samples are combinatorially coupled and multiplied. The DMAS formula is given by:

$$y_{DMAS}(k) = \sum_{i=1}^{M-1} \sum_{j=i+1}^{M} x_i(k - \Delta_i) x_j(k - \Delta_j).$$

(2)

To overcome the dimensionally squared problem of (2), following equations are suggested:[35]

$$\hat{x}_{ij}(k) = \text{sign}[x_i(k - \Delta_i) x_j(k - \Delta_j)] \sqrt{|x_i(k - \Delta_i) x_j(k - \Delta_j)|}.$$

(3)

$$y_{DMAS}(k) = \sum_{i=1}^{M-1} \sum_{j=i+1}^{M} \hat{x}_{ij}(k).$$ (4)

A product in time domain is equivalent to the convolution of the spectra of the signals in the frequency domain. Consequently, new components centered at the zero frequency and the harmonic frequency appear in the spectrum due to the similar ranges of frequency for $x_i(k - \Delta_i)$ and $x_j(k - \Delta_j)$. A band-pass filter is applied on the beamformed output signal to only pass the necessary frequency components, generated after these non-linear operations, while keeping the one centered on $2f_0$ almost unaltered. Finally, the Filtered-DMAS (F-DMAS) is obtained, extensively explained in.[35] The procedure of DMAS algorithm can be considered as a correlation process which uses the auto-correlation of aperture. In other words, the output of this beamformer is based on the spatial coherence of PA signals, and it is a non-linear beamforming algorithm.

## 2.2 Minimum Variance

The output of MV adaptive beamformer is given by:

$$y(k) = \boldsymbol{W}^H(k)\boldsymbol{X}_d(k) = \sum_{i=1}^{M} w_i(k)x_i(k - \Delta_i),$$ (5)

where $\boldsymbol{X}_d(k)$ is time-delayed array detected signals $\boldsymbol{X}_d(k) = [x_1(k), x_2(k), ..., x_M(k)]^T$, $\boldsymbol{W}(k) = [w_1(k), w_2(k), ..., w_M(k)]^T$ is the beamformer weights, and $(.)^T$ and $(.)^H$ represent the transpose and conjugate transpose, respectively. The detected array signals can be written as follows:

$$\boldsymbol{X}(k) = \boldsymbol{s}(k) + \boldsymbol{i}(k) + \boldsymbol{n}(k) = s(k)\boldsymbol{a} + \boldsymbol{i}(k) + \boldsymbol{n}(k),$$ (6)

where $\boldsymbol{s}(k), \boldsymbol{i}(k)$ and $\boldsymbol{n}(k)$ are the desired signal, interference and noise components received by array transducer, respectively. Parameters $s(k)$ and $\boldsymbol{a}$ are the signal waveform and the related steering vector, respectively. MV bemaformer can be used to adaptively weight the calculated samples, and the goal of MV beamformer is to achieve optimal weights in order to estimate the desired signal as accurately as possible. The superiority of the MV algorithm has been evaluated in comparison with static windows, such as Hamming window.[28] To acquire the optimal weights, signal-to-interference-plus-noise ratio ($SINR$) needs to be maximized:[46]

$$SINR = \frac{\sigma_s^2 |\boldsymbol{W}^H \boldsymbol{a}|^2}{\boldsymbol{W}^H \boldsymbol{R}_{i+n} \boldsymbol{W}},\tag{7}$$

where $\boldsymbol{R}_{i+n}$ is the $M \times M$ interference-plus-noise covariance matrix, and $\sigma_s^2$ is the signal power. The maximization of $SINR$ can be gained by minimizing the output interference-plus-noise power while maintaining a distortionless response to the desired signal using following equation:

$$\min_{\boldsymbol{W}} \boldsymbol{W}^H \boldsymbol{R}_{i+n} \boldsymbol{W}, \quad s.t. \quad \boldsymbol{W}^H \boldsymbol{a} = 1.\tag{8}$$

The solution of (8) is given by:[47]

$$\boldsymbol{W}_{opt} = \frac{\boldsymbol{R}_{i+n}^{-1} \boldsymbol{a}}{\boldsymbol{a}^H \boldsymbol{R}_{i+n}^{-1} \boldsymbol{a}}.\tag{9}$$

In practical application, interference-plus-noise covariance matrix is unavailable. Consequently, the sample covariance matrix is used instead of unavailable covariance matrix using N recently received samples and is given by:

$$\hat{\boldsymbol{R}} = \frac{1}{N} \sum_{n=1}^{N} \boldsymbol{X}_d(n) \boldsymbol{X}_d(n)^H.\tag{10}$$

Using MV in medical US imaging encounters some problems which are addressed in,[27] and it is out of this paper discussion, but we briefly review it here. It should be noticed that by applying delays on each element of the array, the steering vector $\boldsymbol{a}$ for each signal waveform becomes a vector of ones.[26–28] The subarray-averaging or spatial-smoothing method can be used to achieve a good estimation of covariance matrix using decorrelation of the coherent signals received by array elements. The covariance matrix estimation using spatial-smoothing can be written as:

$$\hat{\boldsymbol{R}}(k) = \frac{1}{M-L+1} \sum_{l=1}^{M-L+1} \boldsymbol{X}_d^l(k) \boldsymbol{X}_d^l(k)^H,$$ (11)

where $L$ is the subarray length and $\boldsymbol{X}_d^l(k) = [x_d^l(k), x_d^{l+1}(k), ..., x_d^{l+L-1}(k)]$ is the delayed input signal for the $l_{th}$ subarray. Due to limited statistical information, only a few temporal samples are used to estimate covariance matrix. Therefore, to obtain a stable covariance matrix, the diagonal loading ($DL$) technique is used. This method leads to replacing $\hat{\boldsymbol{R}}$ by loaded sample covariance matrix, $\hat{\boldsymbol{R}}_l = \hat{\boldsymbol{R}} + \gamma \boldsymbol{I}$, where $\gamma$ is the loading factor given by:

$$\gamma = \Delta.trace\{\hat{\boldsymbol{R}}(k)\},$$ (12)

where $\Delta$ is a constant related to subarray length. Also, temporal averaging method can be applied along with spatial averaging to gain the resolution enhancement while the contrast is retained. The estimation of covariance matrix using both temporal averaging and spatial smoothing in given by:

$$\hat{\boldsymbol{R}}(k) = \frac{1}{(2K+1)(M-L+1)} \times \sum_{n=-K}^{K} \sum_{l=1}^{M-L+1} \boldsymbol{X}_d^l(k+n) \boldsymbol{X}_d^l(k+n)^H,$$ (13)

where temporal averaging is used over $(2K + 1)$ samples. After estimation of covariance matrix, optimal weights are calculated by (9) and (13) and finally the output of MV beamformer is given by:

$$\hat{y}(k) = \frac{1}{M - L + 1} \sum_{l=1}^{M-L+1} \boldsymbol{W}_*^H(k) \boldsymbol{X}_d^l(k). \tag{14}$$

where $\boldsymbol{W}_*(k) = [w_1(k), w_2(k), ..., w_L(k)]^T$.

## 3 Proposed Method

In this paper, it is proposed to use the MV adaptive beamformer instead of the existing DAS algebra inside DMAS mathematical expansion. To illustrate this, consider the expansion of the DMAS algorithm which can be written as follows:

$$
\begin{aligned}
y_{DMAS}(k) = \sum_{i=1}^{M-1} \sum_{j=i+1}^{M} x_{id}(k) x_{jd}(k) = \\
\Big[ x_{1d}(k) x_{2d}(k) + x_{1d}(k) x_{3d}(k) + ... + x_{1d}(k) x_{Md}(k)) \Big] \\
+ \Big[ x_{2d}(k) x_{3d}(k) + x_{2d}(k) x_{4d}(k) + ... + x_{2d}(k) x_{Md}(k) \Big] + ... \\
+ \Big[ x_{(M-2)d}(k) x_{(M-1)d}(k) + x_{(M-2)d}(k) x_{Md}(k) \Big] + \Big[ x_{(M-1)d}(k) x_{Md}(k) \Big],
\end{aligned} \tag{15}
$$

where $x_{id}(k)$ and $x_{jd}(k)$ are delayed detected signals for element $i$ and $j$, respectively, and we hold this notation all over this section. As can be seen, there is a DAS in every terms of the expansion, and it can be used to modify the DMAS beamformer. To illustrate this, consider the

following equation:

$$y_{DMAS}(k) = \sum_{i=1}^{M-1} \sum_{j=i+1}^{M} x_{id}(k) x_{jd}(k) =$$

$$x_{1d}(k) \underbrace{\left[ x_{2d}(k) + x_{3d}(k) + x_{4d}(k) + ... + x_{Md}(k) \right]}_{\text{first term}} + x_{2d}(k) \underbrace{\left[ x_{3d}(k) + x_{4d}(k) + ... + x_{Md}(k) \right]}_{\text{second term}}$$

$$+ ... + x_{(M-2)d}(k) \underbrace{\left[ x_{(M-1)d}(k) + x_{Md}(k) \right]}_{\text{(M-2)th term}} + \underbrace{\left[ x_{(M-1)d}(k).x_{Md}(k) \right]}_{\text{(M-1)th term}}.$$

$$(16)$$

In (16), in every terms, there exists a summation procedure which is a type of DAS algorithm. It is proposed to use MV adaptive beamformer for each term instead of DAS. In other words, since DAS is a non-adaptive beamformer and considers all calculated samples for each element of the array the same as each other, consequently, the acquired image by each term is a low quality image with high levels of sidelobes and broad mainlobe. In order to use MV instead of every DAS in the expansion, we need to carry out some modifications and prepare the expansion in (16) for the proposed method. Following section contains the essential modifications.

### 3.1 Modified DMAS

It should be noticed that the quality of covariance matrix estimation in MV highly depends on the selected length of subarray. The upper boundary is limited to $M/2$ and the lower boundary to 1. Choosing $L = M/2$ leads to resolution enhancement at the cost of robustness, and $L = 1$ leads to resolution reduction and robustness increment. In (16), each term can be considered as a DAS algorithm with different number of elements of array. In other words, the number of samples of elements contributing in the existing DAS is different in each term, which results from the nature of DMAS algorithm. To illustrate this, consider the length of array $M$ and $L = M/2$. There will be $M-1$ terms in DMAS expansion, while first term contains $M-1$ entries, second term contains $M-2$ entries and finally the last term contains only one entry. Limited number of entries in each term causes problem for MV algorithm due to the limited length of the subarray. This problem can be addressed by adding the unavailable elements in each term in order to acquire large enough number of available elements and consequently high quality covariance matrix estimation. The extra terms, needed to address the problem, are given by:

$$
\begin{aligned}
y_{extra}(k) &= \sum_{i=M-2}^{2} \sum_{j=i-1}^{1} x_{id}(k) x_{jd}(k) + y_{extra^*} \\
&= x_{(M-2)d}(k) \Big[ x_{(M-3)d}(k) + x_{(M-4)d}(k) + ... + x_{2d}(k) + x_{1d}(k) \Big] \\
&+ x_{(M-3)d}(k). \Big[ x_{(M-4)d}(k) + x_{(M-5)d}(k) + ... + x_{2d}(k) ] + x_{1d}(k) \Big] \\
&+ ... + x_{3d}(k). \Big[ x_{2d}(k) + x_{1d}(k) \Big] + x_{2d}(k) x_{1d}(k) + y_{extra^*}(k)
\end{aligned}
\tag{17}
$$

,

where

$$
y_{extra^*}(k) = x_{Md}(k) \Big[ x_{(M-1)d}(k) + x_{(M-2)d}(k) + ... + x_{3d}(k) + x_{2d}(k) + x_{1d}(k) \Big].
\tag{18}
$$

(17) is used to make the terms in (16) ready to adopt a MV algorithm. Finally, by adding (16) and (17), a modified version of DMAS algorithm namely modified DMAS (MDMAS) is obtained and

can be written as:

$$y_{MDMAS}(k) = y_{DMAS}(k) + y_{extra}(k)$$

$$= \sum_{i=1}^{M} \sum_{j=1, j \neq i}^{M} x_{id}(k)x_{jd}(k) =$$

$$= x_{1d}(k) \underbrace{\left[x_{2d}(k) + x_{3d}(k) + ... + x_{(M-1)d}(k) + x_{Md}(k)\right]}_{\text{first term}}$$

$$+ x_{2d}(k) \underbrace{\left[x_{1d}(k) + x_{3d}(k) + ... + x_{(M-1)d}(k) + x_{Md}(k)\right]}_{\text{second term}} \qquad (19)$$

$$+ ... + x_{(M-1)d}(k). \underbrace{\left[x_{1d}(k) + x_{2d}(k) + ... + x_{(M-2)d}(k) + x_{Md}(k)\right]}_{\text{first term}}$$

$$+ x_{Md}(k) \underbrace{\left[x_{1d}(k) + x_{2d}(k) + ... + x_{(M-2)d}(k) + x_{(M-1)d}(k)\right]}_{\text{Mth term}}.$$

The introduced algorithm in (19) has been evaluated by simulations, and it is proved that this formula can be a modification of DMAS algebra with the same results. To put it more simply, (19) is the multiplication of DMAS output by 2, and since all the cross-products are considered twice, simulations give the same results. Now, the combination of MDMAS algorithm and MV beamformer is mathematically satisfying and instead of every terms in (19), MV can be implemented using all entities in each term. The expansion of MDMAS combined with MV beamformer can be

written as follows:

$$y_{MV-DMAS}(k) = \sum_{i=1}^{M} x_{id}(k)\big(\boldsymbol{W}_{i,M-1}^{H}(k)\boldsymbol{X}_{id,M-1}(k)\big) =$$

$$\sum_{i=1}^{M} x_{id}(k)\bigg(\sum_{j=1,j\neq i}^{M} w_j(k)x_{jd}(k)\bigg) = \sum_{i=1}^{M} x_{id}(k)\bigg(\boldsymbol{W}^{H}(k)\boldsymbol{X}_d(k) - w_i(k)x_{id}(k)\bigg) =$$

$$\sum_{i=1}^{M} x_{id}(k)\bigg(\sum_{j=1}^{M} w_j(k)x_{jd}(k) - w_i(k)x_{id}(k)\bigg) = \tag{20}$$

$$\sum_{i=1}^{M} x_{id}(k)\underbrace{\bigg(\sum_{j=1}^{M} w_j(k)x_{jd}(k)\bigg)}_{MV} - \sum_{i=1}^{M} x_{id}(k)\bigg(w_i(k)x_{id}(k)\bigg),$$

where $\boldsymbol{W}_{i,M-1}$ and $\boldsymbol{X}_{id,M-1}$ are almost the same as $\boldsymbol{W}(k)$ and $\boldsymbol{X}_d(k)$ used in (5), respectively, but the $i_{th}$ element of the array is ignored in calculation and as a result, the length of these vectors becomes $M-1$ instead of $M$. Considering (20), the expansion can be written based on a summation which is considered as a DAS algebra. To illustrate this, consider following expansion:

$$y_{MV-DMAS}(k) = \sum_{i=1}^{M} x_{id}(k)\underbrace{\bigg(\sum_{j=1}^{M} w_j(k)x_{jd}(k)\bigg)}_{MV} - \sum_{i=1}^{M} x_{id}(k)\bigg(w_i(k)x_{id}(k)\bigg) =$$

$$\underbrace{\sum_{i=1}^{M} x_{id}(k)\underbrace{\bigg(\sum_{j=1}^{M} w_j(k)x_{jd}(k)\bigg)}_{MV} - w_i(k)x_{id}^2(k)}_{i_{th}term} . \tag{21}$$

It is proved that DAS leads to low quality images and high levels of sidelobe and obviously in (21), expansion leads to a summation and this summation can be considered as a DAS. As a final step of MVB-DMAS development, it is proposed to use another MV instead of DAS in order to reduce the contribution of off-axis signals and noise of imaging system. To put it more simply, considering (21), each term is contributed in a summation process which is regarded as a DAS, represented in (1). Since (5) leads to image enhancement compared to (1), it is expected to improve the image quality in terms of resolution and levels of sidelobe having MV instead of outer summation in (21). MVB-DMAS formula can be written as follows:

$$y_{MVB-DMAS}(k) = \sum_{i=1}^{M} w_{i,new} \underbrace{\left( x_{id}(k) \left( \sum_{j=1}^{M} w_j(k) x_{jd}(k) \right) - w_i(k) x_{id}^2(k) \right)}_{i_{th} term},$$
(22)

where $w_{i,new}$ is the calculated weight for each term in (22) using (9) while the steering vector is a vector of ones. It should be noticed that when there is a multiplication, resulting in squared dimension, the method mentioned in (4) is used to prevent the squared dimension. Moreover, there are two MV algorithms inside the proposed method, one on the delayed signals and one on the $i_{th}$ term obtained with (21). Since we face with the correlation procedure of DMAS, including product function in time domain, in the proposed method, necessary band-pass filter is applied in (21) for each term, before outer summation. In other words, each term in the proposed method in (22) is filtered to only pass the necessary components, generated after the non-linear operations, and then all of them are contributed in the second MV algorithm. In the section 4, it is shown that MVB-DMAS beamformer results in resolution improvement and sidelobes level reduction.

## 4 NUMERICAL RESULTS AND PERFORMANCE ASSESSMENT

In this section, numerical results are presented to illustrate the performance of the proposed algorithm in comparison with DAS, DMAS and MV.

### 4.1 Simulated Point Target

#### 4.1.1 Simulation Setup

The K-wave Matlab toolbox was used to simulate the numerical study.[48] Eleven 0.1 $mm$ radius spherical absorbers as initial pressure were positioned along the vertical axis every 5 $mm$ beginning 25 $mm$ from the transducer surface. The imaging region was 20 $mm$ in lateral axis and 80 $mm$ in vertical axis. A linear-array having $M$=128 elements operating at 5 $MHz$ central frequency and 77% fractional bandwidth was used to detect the PA signals generated from the defined initial pressures. The speed of sound was assumed to be 1540 $m/s$ during simulations. The sampling frequency was 50 $MHz$, subarray length $L$=$M$/2, $K$=5 and $\Delta = 1/100L$ for all the simulations. Also, a band-pass filter was applied by a Tukey window ($\alpha$=0.5) to the beamformed signal spectra, covering 6-16 $MHz$, to pass the necessary information.

#### 4.1.2 Qualitative Evaluation

Fig. 1(a), Fig. 1(b), Fig. 1(c) and Fig. 1(d) show the output of DAS, MV, DMAS and MVB-DMAS beamformers, respectively. It is clear that DAS and DMAS result in low resolution images, and at the high depths of imaging both algorithms lead to a wide mainlobe. However, DMAS leads to lower level of sidelobes and a higher resolution. In Fig. 1(b), it can be seen that MV results in high resolution, but the high level of sidelobes affect the reconstructed image. Formed image using MVB-DMAS is shown in Fig. 1(d) where the resolution of MV beamformer is maintained and the level of sidelobe are highly degraded compared to MV. To assess the different beamforming algorithms in details, lateral variations of the formed images are shown in Fig. 2. Lateral variations at the depth of 50 $mm$ is shown in Fig. 2(c) where DAS, MV, DMAS and MVB-DMAS result in about -40 $dB$, -50 $dB$, -55 $dB$ and -65 $dB$ sidelobes, respectively. On the other hand, width of

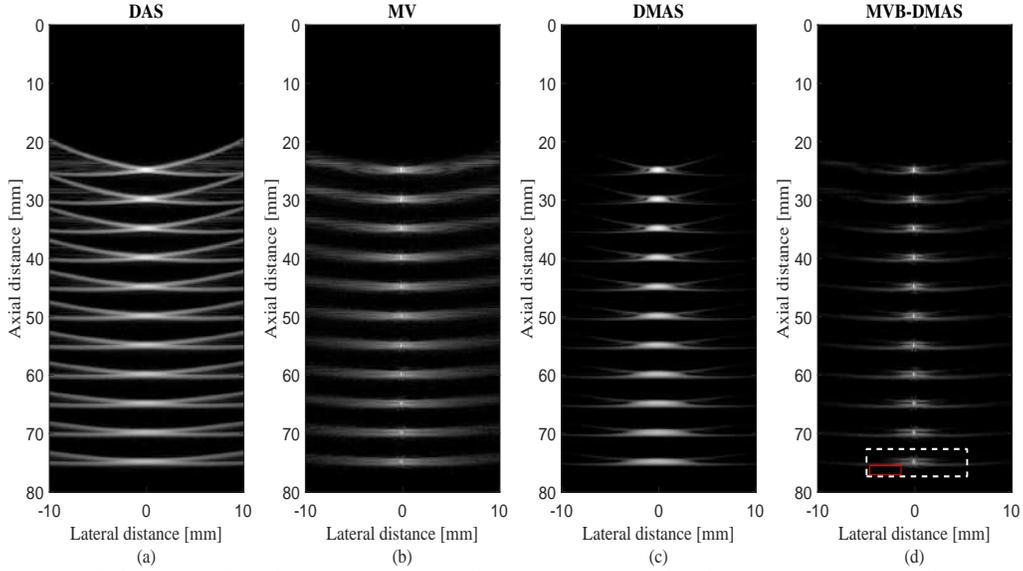

Fig 1: Images of the simulated point targets phantom using the linear-array transducer. (a) DAS, (b) MV, (c) DMAS, (d) MVB-DMAS. All images are shown with a dynamic range of 60 $dB$. Noise was added to the detected signals considering a SNR of 50 $dB$.

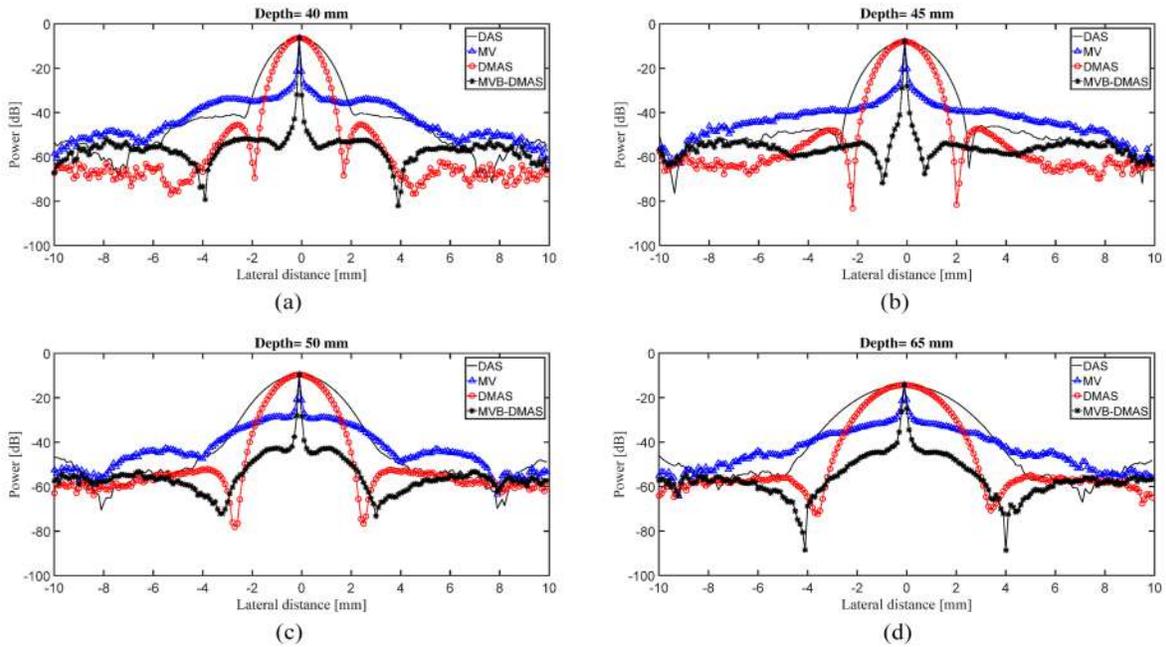

Fig 2: Lateral variations of DAS, MV, DMAS and MVB-DMAS at the depths of (a) 40 $mm$, (b) 45 $mm$, (c) 50 $mm$ and (d) 65 $mm$.

mainlobe can be regarded as a parameter, indicating the resolution metric. It can be seen that MV and MVB-DMAS result in significant higher resolution in comparison with DAS and DMAS.

Table 1: FWHM ($\mu m$) values (in -6 $dB$) at the different depths.

| Depth(mm) \ Beamformer | DAS | DMAS | MV | MVB-DMAS |
|---|---|---|---|---|
| 25 | 1200 | 850 | 93 | 54 |
| 30 | 1476 | 1059 | 99 | 54 |
| 35 | 1842 | 1286 | 115 | 61 |
| 40 | 2277 | 1584 | 133 | 72 |
| 45 | 2710 | 1862 | 143 | 82 |
| 50 | 3565 | 2355 | 172 | 95 |
| 55 | 3800 | 2535 | 187 | 102 |
| 60 | 4400 | 2937 | 226 | 113 |
| 65 | 4967 | 3273 | 288 | 123 |
| 70 | 5512 | 3639 | 305 | 146 |
| 75 | 6625 | 4244 | 471 | 212 |

Table 2: SNR ($dB$) values at the different depths.

| Depth(mm) \ Beamformer | DAS | DMAS | MV | MVB-DMAS |
|---|---|---|---|---|
| 25 | 30.7 | 51.1 | 42.7 | 57.7 |
| 30 | 28.6 | 46.9 | 39.3 | 53.2 |
| 35 | 26.8 | 43.6 | 37.7 | 50.8 |
| 40 | 25.2 | 40.2 | 35.3 | 48.2 |
| 45 | 23.5 | 36.2 | 34.4 | 46.5 |
| 50 | 21.7 | 32.0 | 33.2 | 45.0 |
| 55 | 20.1 | 28.2 | 32.4 | 50.8 |
| 60 | 18.5 | 25.0 | 31.3 | 48.2 |
| 65 | 17.2 | 22.6 | 30.4 | 46.5 |
| 70 | 16.3 | 20.7 | 30.1 | 45.0 |
| 75 | 15.5 | 19.1 | 29.0 | 43.3 |

### 4.1.3 Quantitative Evaluation

To quantitatively compare the performance of the beamformers, the full-width-half-maximum (FWHM) in -6 $dB$ and signal-to-noise ratio (SNR) are calculated in all imaging depths using point targets in the reconstructed images. The results for FWHM and SNR are shown in TABLE 1 and TABLE 2, respectively. As can be seen in TABLE 1, MVB-DMAS results in the narrowest -6 $dB$ width of mainlobe in all imaging depths compared to other beamformers. In particular, consider depth of 50 $mm$ where FWHM for DAS, DMAS, MV and MVB-DMAS is about 3565 $\mu m$, 2355 $\mu m$, 172 $\mu m$ and 95 $\mu m$, respectively. More importantly, the FWHM differentiation of the first and last imaging depth indicates that MVB-DMAS and MV techniques variate 158 $\mu m$ and 378 $\mu m$, respectively, while DAS and DMAS variate 5425 $\mu m$ and 3394 $\mu m$, respectively. As a result, FWHM is more stabilized using MVB-DMAS and MV in comparison with DAS and DMAS. The represented SNRs in TABLE 2 are calculated using following equation:

$$SNR = 20 \log_{10} P_{signal}/P_{noise}. \tag{23}$$

where $P_{signal}$ and $P_{noise}$ are difference of maximum and minimum intensity of a rectangular region including a point target (white dashed rectangle in Fig. 1(d)), and standard deviation of the noisy part of the region (red rectangle in Fig. 1(d)), respectively.[39,49] As can be seen in TABLE 2, MVB-DMAS outperforms other beamformers in SNR. Consider, in particular, the depth of 50 $mm$ where SNR for DAS, DMAS, MV and MVB-DMAS is 21.7 $dB$, 32.0 $dB$, 33.2 $dB$ and 45.0 $dB$, respectively.

### 4.2 Sensitivity to Sound Velocity Inhomogeneities

In this section, the proposed method is evaluated in the term of robustness against the sound velocity errors resulting from medium inhomogeneities which are inevitable in practical imaging. The simulation design for Fig. 1 is used in order to investigate the robustness, except that the sound velocity is overestimated by 5%, which covers and may be more than the typical estimation error.[26,27] It should be noticed that in the simulation, we have intentionally overestimated the sound velocity by 5% to evaluate the proposed method. This is important since we usually face this phenomenon in the practical situations where the medium is inhomogeneous, but the images are reconstructed assuming a sound velocity of 1540 $m/s$. As can be seen in Fig. 3(b), MV leads to higher resolution compared to DAS, but the high levels of sidelobe and negative effects of overestimated sound velocity still affect the reconstructed image. It should be noticed that the appeared noise and artifacts in Fig. 3(b) are due to the overestimated sound velocity and the temporal averaging with K=5. DMAS, in Fig. 3(c), reduces these negative effects, but the resolution is not well enough. As can be seen in Fig. 3(d), MVB-DMAS results in the high negative effects reduction of DMAS and the high resolution of MV. However, the reconstructed image using MVB-DMAS contains more artifacts compared to DMAS, which is mainly as a result of the lower SNR of MVB-DMAS compared to DMAS. Fig. 4 shows the lateral variation of the reconstructed images in Fig. 3. As can be seen, MVB-DMAS detects the peak amplitude of point target as well as DAS. The resolution of the formed image using MVB-DMAS is improved in comparison with DAS and DMAS. Moreover, the levels of sidelobe using MVB-DMAS is reduced in comparison with other mentioned beamformers.

### 4.3  *Effects of Varying L*

To evaluate the effects of varying L, the proposed method has been implemented using $L$=64, $L$=45, $L$=32 and $L$=16. The lateral variations of the formed images at the depth of 45 $mm$ are presented in Fig. 5. Clearly, increasing $L$ results in a higher resolution, and lower level of sidelobes. Moreover, the SNR, in two imaging depths, is presented in Table 3. It is shown that SNR does not significantly vary for different $L$. However, $L$=45 results in higher SNR. In addition, Table 4 shows the calculated FWHM for different amounts of $L$, and it proves that FWHM is reduced with higher $L$.

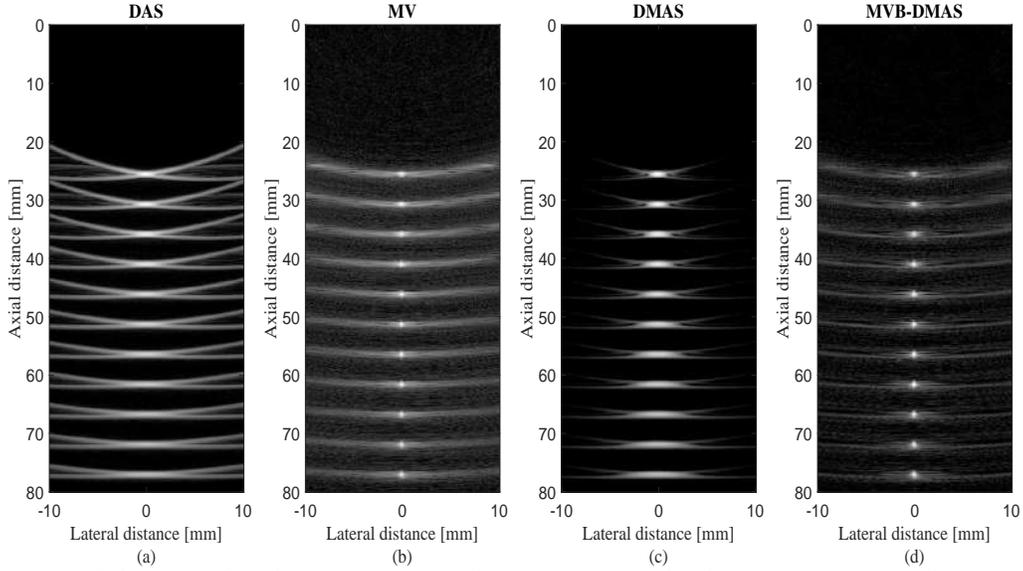

Fig 3: Images of the simulated point targets phantom using the linear-array transducer. (a) DAS, (b) MV, (c) DMAS, (d) MVB-DMAS. All images are shown with a dynamic range of 60 $dB$. The sound velocity is overestimated about 5 %. Noise was added to the detected signals considering a SNR of 50 $dB$.

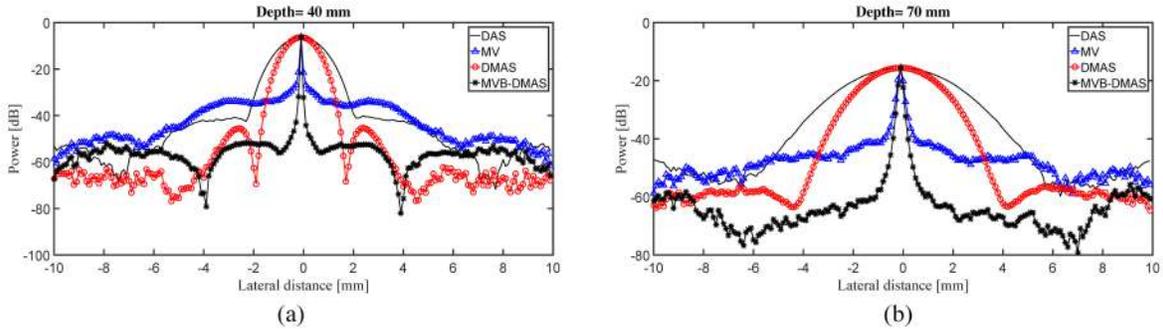

Fig 4: Lateral variations of DAS, MV, DMAS and MVB-DMAS at the depths of (a) 45 $mm$ and (b) 70 $mm$ while the sound velocity is 5% overestimated.

### 4.4 Effects of Coherence Weighting

The proposed algorithm is also evaluated when it has been combined with CF. To have a fair comparison, all the beamformers are combined with CF weighting. The reconstructed images along with the corresponding lateral variations are shown in Fig. 6 and Fig. 7, respectively. As can be seen in Fig. 6, MVB-DMAS+CF results in higher resolution and lower sidelobes compared to other beamformers. The higher resolution of MVB-DMAS+CF is visible compared to DMAS+CF, especially at high depths of imaging. In addition, it is clear that MVB-DMAS+CF reduces the sidelobes compared to MV+CF and improves the target detectability. To have a better comparison, consider Fig. 7 where MVB-DMAS+CF outperforms other beamformers in the terms of width of mainlobe in -6 $dB$ and sidelobes. It is shown that sidelobes for DAS, DMAS, MV and MVB-DMAS, when all of them are combined with CF, are about -120 dB, -120 dB, -111 dB and -160 dB, respectively, showing the superiority of MVB-DMAS+CF.

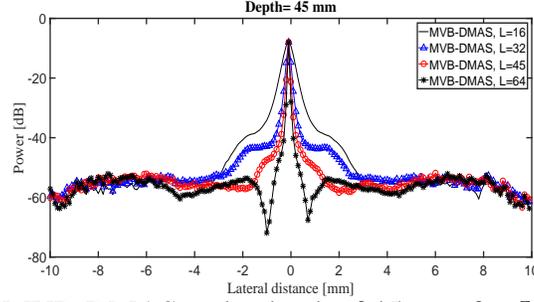

Fig 5: Lateral variations of MVB-DMAS at the depth of 45 $mm$ for $L$=16, $L$=32 , $L$=45 and $L$=64.

Table 3: SNR ($dB$) values of MVB-DMAS for the different amounts of $L$.

| Number of $L$ Depth($mm$) | 16 | 32 | 45 | 64 |
|---|---|---|---|---|
| 45 | 67.5 | 67.2 | 67.3 | 66.0 |
| 65 | 61.8 | 61.6 | 62.4 | 61.0 |

## 5 EXPERIMENTAL RESULTS

To evaluate the MVB-DMAS algorithm, in this section results of designed experiments are presented.

### 5.1 Experimental Setup

A linear-array of PAI system was used to detect the PA waves and the major components of system include an ultrasound data acquisition system, Vantage 128 Verasonics (Verasonics, Inc., Redmond, WA), a Q- switched Nd:YAG laser (EverGreen Laser, Double-pulse Nd: YAG system) with

Table 4: FWHM ($\mu m$) values of MVB-DMAS (in -6 $dB$) for the different amounts of $L$.

| Number of $L$<br>Depth($mm$) | 16 | 32 | 45 | 64 |
|---|---|---|---|---|
| 45 | 480 | 199 | 95 | 59 |
| 65 | 735 | 259 | 161 | 97 |

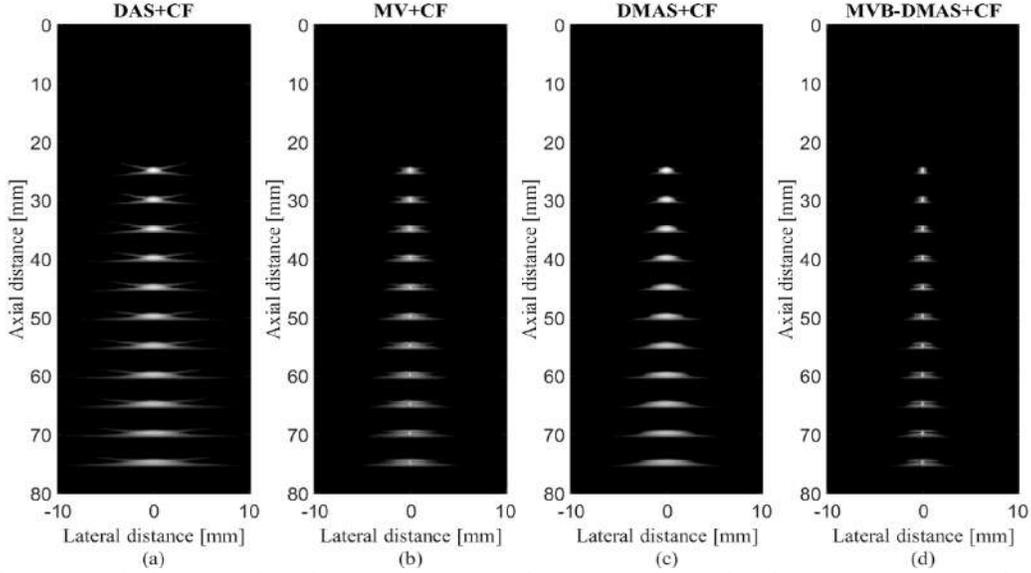

Fig 6: Images of the simulated point targets phantom using the linear-array transducer. (a) DAS+CF, (b) MV+CF, (c) DMAS+CF, (d) MVB-DMAS+CF. All images are shown with a dynamic range of 60 $dB$. Noise was added to the detected signals considering a SNR of 50 $dB$.

a pulse repetition rate of 25 $Hz$, wavelength 532 $nm$ and a pulse width of 10 $ns$. A transducer array (L7-4, Philips Healthcare) with 128 elements and 5.2 $MHz$ central frequency was used as a receiver. A function generator is used to synchronize all operations (i.e., laser firings and PA signal recording). The data sampling rate was 20.8320 $MHz$. The schematic of the designed system is presented in Fig. 8, and a gelatin-based phantom used as imaging target is shown in Fig. 9, including two blood inclusions to provide optoacoustic properties. The experimental setup for PA linear-array imaging is shown in Fig. 10 where two parallel wire are used as phantom for another experiment. It should be noticed that in all experiments, surface of the transducer is perpendicular to the imaging targets. Thus, it is expected to see a cross section of the targets. A band-pass filter was applied by a Tukey window ($\alpha$=0.5) to the beamformed signal spectra, covering 6-13 $MHz$,

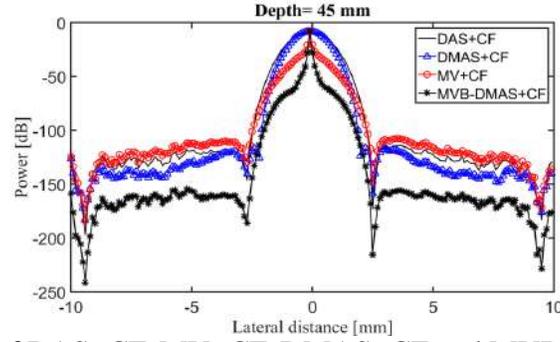

Fig 7: Lateral variations of DAS+CF, MV+CF, DMAS+CF and MVB-DMAS+CF at the depths of 45 $mm$.

to pass the necessary information.

### 5.2 Qualitative Evaluation

The reconstructed images using the phantom shown in Fig. 9 are presented in Fig. 11. Clearly, there are three structures seen in the reconstructed images, Fig. 11, which two of them are blood inclusions, and the first one is because of the small fracture on the upper part of the phantom shown in Fig. 9. As is demonstrated, DAS leads to a low resolution image having high level of sidelobe, especially the target at the depth of 35 $mm$. MV leads to a higher resolution in comparison with DAS, but negative effects of the high level of sidelobes are obvious Fig. 11(b), and the background of the reconstructed image are affected by noise. DMAS enhances the image in the terms of sidelobes and artifacts, but still provides a low resolution image. MVB-DMAS leads to a higher resolution image having lower sidelobes compared to DAS, DMAS and MV. It

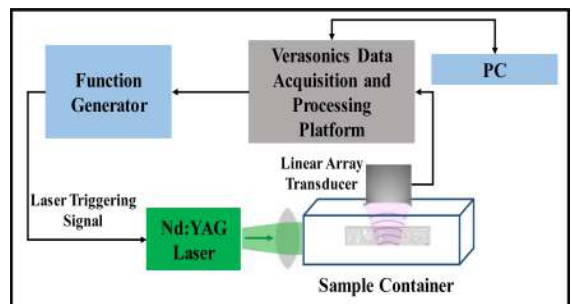

Fig 8: Schematic of the experimental setup.

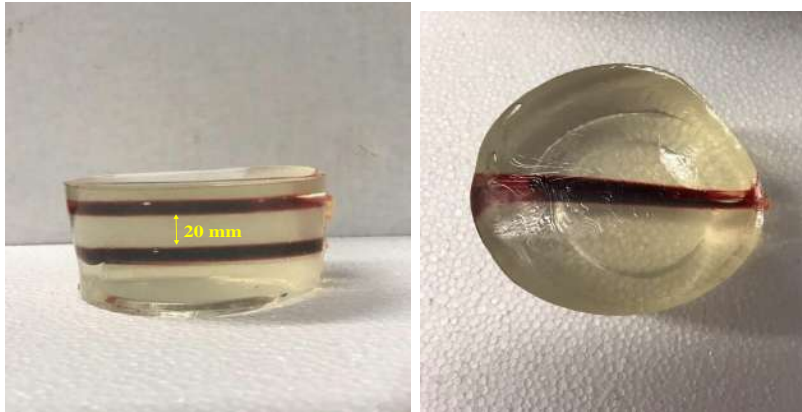

Fig 9: Photographs of the phantom used in the first experiment.

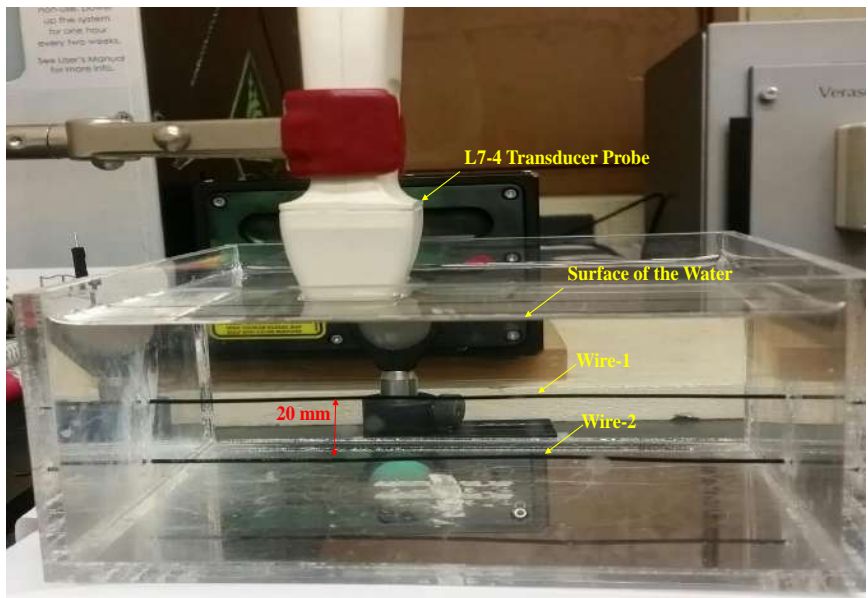

Fig 10: Experimental setup of PA linear-array imaging of two parallel wires.

is clear that MVB-DMAS provides the high resolution of MV and low sidelobes of DMAS. The line above the targets is due to the PA signal generation at the surface of the phantom (due to the top illumination) and is supposed to look like what is seen in Fig. 11(d), considering the area of illumination, its location and laser beam profile. The phantom we used was a bit old, and its surface was slightly dried. Hence, when we added the ultrasound gel on the top of the sample, there was a rather large impedance mismatch created at the interface between the surface of the sample and the US gel. In our other similar tests (Fig. 12), we observed the similar artifact, but much weaker. In

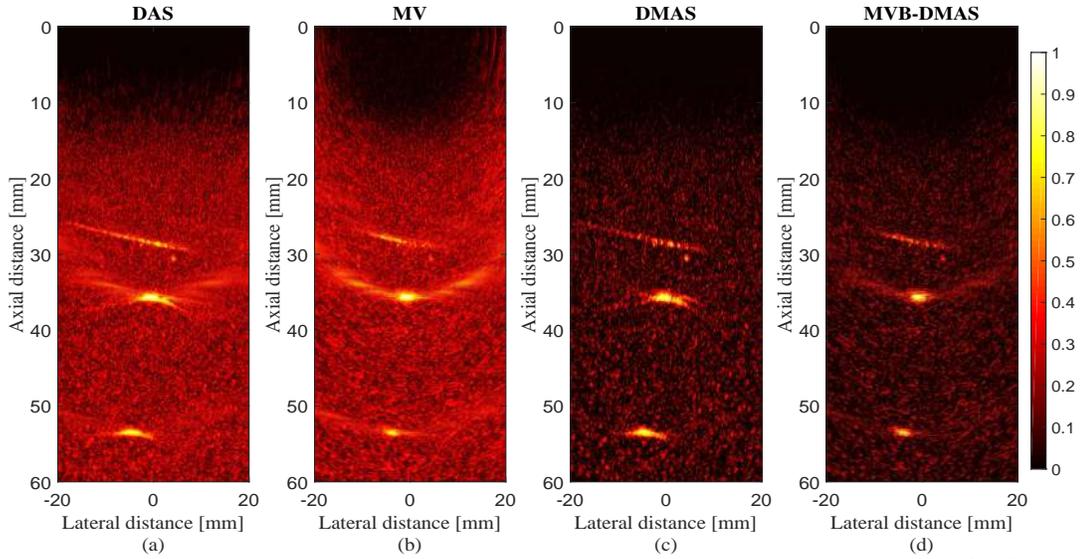

Fig 11: Images of the phantom shown in Fig. 9 using the linear-array transducer. (a) DAS, (b) MV, (c) DMAS and (d) MVB-DMAS. All images are shown with a dynamic range of 60 $dB$.

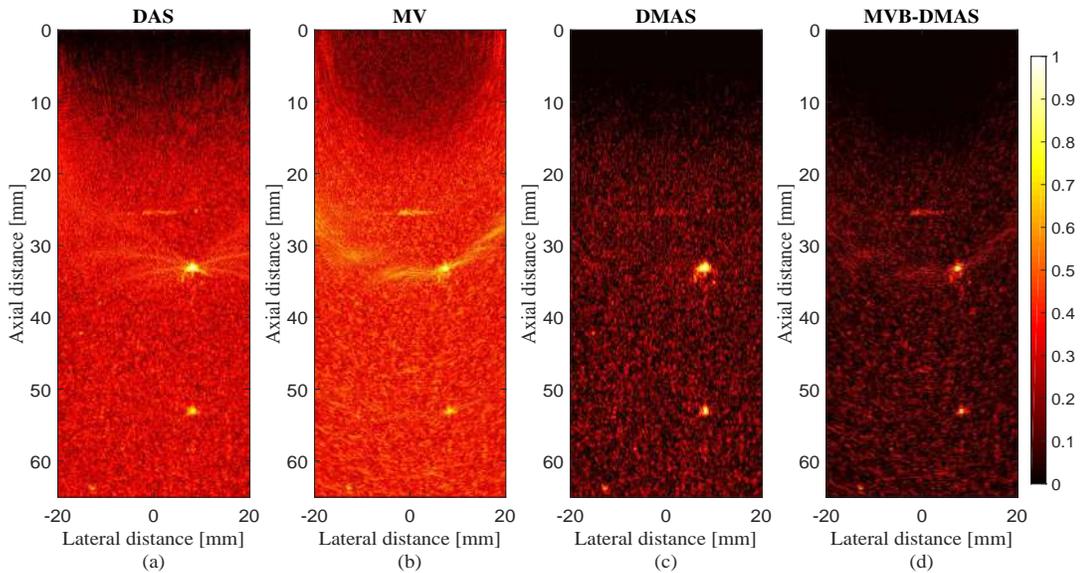

Fig 12: Images of the wires shown in Fig. 10 using the linear-array transducer. (a) DAS, (b) MV, (c) DMAS and (d) MVB-DMAS. All images are shown with a dynamic range of 60 $dB$.

Fig. 11(a) and Fig. 11(c), the extended version of the line is seen which is considered as an artifact.

DAS and DMAS stretch imaging targets (please see the two imaging targets in Fig. 11(a) and Fig. 11(c)). On the other hand, the stretch in MVB-DMAS is much less due to the correlation process and two stages of MV. The reconstructed images for the designed experiment shown in Fig. 10 are shown in Fig. 12. Since the surface of the transducer is perpendicular to the wires, it is expected to see the targets like points. As is demonstrated in Fig. 12(a), DAS results in low resolution points, along with high levels of artifacts, especially at the depth of about 30 $mm$. In Fig. 12(b), MV leads to resolution improvement while the image is still suffers from high level of sidelobes. The reconstructed image using DMAS, shown in Fig. 12(c), contains low level of sidelobes, but the resolution is low. Finally, MVB-DMAS provides an image with characteristics of DMAS and MV, which are reduced sidelobes and high resolution, respectively. Fig. 13 demonstrates the lateral variations of the beamformers at the depth of 33 $mm$ of Fig. 12. As shown in the green circle, MVB-DMAS results in a narrower width of mainlobe and lower sidelobes (see the arrows)

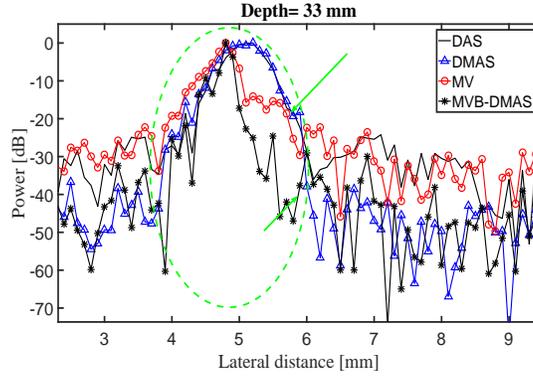

Fig 13: Lateral variations at the depth of 33 $mm$.

### 5.3 Quantitative Evaluation

To compare the experimental results quantitatively, SNR and Contrast Ratio (CR) metrics are used. TABLE 5 and TABLE 6 show the calculated SNR and CR for the two targets in the Fig. 11. CR formula is explained in.[35] As can be seen, the calculated metrics show that MVB-DMAS outperforms other beamformers. In other words, it leads to higher SNR and CR.

Table 5: SNR ($dB$) values at the different depths using the targets in Fig. 11.

| Depth($mm$) \ Beamformer | DAS | DMAS | MV | MVB-DMAS |
|---|---|---|---|---|
| 35 | 28.6 | 46.96 | 39.32 | 53.21 |
| 55 | 24.20 | 45.87 | 37.91 | 50.11 |

Table 6: CR ($dB$) values at the different depths using the targets in Fig. 11.

| Depth($mm$) \ Beamformer | DAS | DMAS | MV | MVB-DMAS |
|---|---|---|---|---|
| 35 | 30.4 | 33.8 | 27 | 36.3 |
| 55 | 26.3 | 30.1 | 26.2 | 32.5 |

### 5.4 In Vivo Imaging

We imaged median antebrachial vein of a 30 years old Middle Eastern male in vivo (see Fig. 14). The illumination was done from side using a single large diameter PMMA fiber (10 $mm$). L22-14$v$ transducer was used to collect the PA signals while it was positioned perpendicular to the vein. The institutional review board at Wayne State University (Independent Investigational Review Board, Detroit, MI) approved the study protocol, and informed consent was obtained from the individual before enrolment in the study. In the reconstructed images, as shown in Fig. 15, the top and bottom of the antebrachial vein showed up. It can be seen that the reconstructed image using MVB-DMAS has lower sidelobes, noise and artifacts compared to other methods, and the cross-sections (top and bottom) of the vein are more detectable.

## 6 Discussion

The main improvement gained by the introduced method is that the high resolution of MV beamforming algorithm is retained while the level of sidelobes are reduced. PA images reconstructed by DAS bemaformer have a low quality, along with high effects of off-axis signals and high sidelobes. This is mainly due to the blindness of DAS. In fact, the DAS algorithm is a procedure in which all contributing samples are treated identically. On the other hand, DMAS beamformer is a non-linear algorithm and leads to high level of off-axis signals rejection due to its correlation process. In DMAS beamformer, all the calculated samples are weighted using a linear combination of the received signals. This procedure makes DMAS a non-blind beamforming algorithm which results in lower effects of off-axis signals and higher contrast reconstructed images compared to DAS. However, the resolution improvement by DMAS is not good enough in comparison with the MV algorithm. In MV beamformer, samples are weighted adaptively resulting significant resolution improvement. However, it leads to high level of sidelobes. Therefore, we face two types of beam-formers which one of them (DMAS) results in sidelobes improvement, and the other one (MV) leads to significant resolution enhancement.

Table 7: Computational operation and processing time(s)

| Metric \ Beamformer | DAS | DMAS | MV | MVB-DMAS |
|---|---|---|---|---|
| Order | $M$ | $M^2$ | $M^3$ | $M^3$ |
| Processing Time | 1.1 | 10.8 | 90.9 | 187.1 |

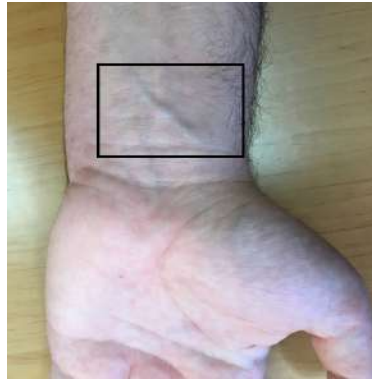

Fig 14: The photo of the median antebrachial vein of a 30 years old Middle Eastern male.

The expansion of DMAS algebra shows there are multiple terms which each of them can be interpreted as a DAS with different lengths of array. This could be the source of the low resolution of DMAS algorithm, and using MV instead of these terms can be an appropriate choice to improve the resolution. However, as shown in (16), the number of contributing samples in each term of the expansion is different. The length of the subarray in the spatial smoothing highly effects the performance of MV algorithm, and in (16) there are some terms representing a low length of array and subarray. To address this problem, necessary terms are added to each term, and then MV algorithm is applied on it. The superiority of MV has been proved compared to DAS, and it is expected to have resolution improvement using MV instead of the existing DAS inside the expansion. This method has been used in the introduced algorithm twice to suppress the artifacts and sidelobes of MV. In other words, there are two MV algorithms inside the proposed method, one on the delayed signals and one on the $i_{th}$ term of (21). The MV implemented on the delayed signals improves the resolution, but since there is another summation procedure interpreting as DAS, shown in 16, the level of sidelobes and artifacts reduce the image quality. Second MV is implemented on the $i_{th}$ term of (21) to use the properties of MV algorithm in order to improve the image quality. It should be noticed that since the expansion of DMAS is used to integrate the MV algorithm for resolution improvement, there are multiplication operations in the introduced algorithm. The same as DMAS, a band-pass filter is needed to only pass the necessary information.[35] The proposed algorithm adaptively calculates the weights for each samples, which improves the resolution. Since the correlation procedure of DMAS contributes in the proposed method, the sidelobe level of MV are reduced while the resolution is retained due to the existence of MV in the proposed method. MVB-DMAS has been evaluated numerically and experimentally. It should be noticed that the processing time of the proposed method is higher than other mentioned beamformers. TABLE 7 shows the order of beamformers computations and corresponding processing time. The correlation process of DMAS needs more time compared to DAS, and MV needs time to adaptive calculation of the weights. MVB-DMAS uses two stages of MV algorithm and a correlation procedure, so it is expected to result in higher processing time compared to MV and DMAS. The computational complexity for calculating the weighting coefficients in MVB-DMAS is in the order of $O(L^3)$. Considering the fact that $L$ supposed to be a fraction of $M$, the computational complexity is a function of $M^3$. Given the weighting coefficient, the computational complexity of the reconstruction procedure is a function of $M$, so the bottle neck of the computational burden is $M^3$, which is the same as regular MV algorithm. Note that, the complexity of DMAS and DAS are $O(M^2)$ and $O(M)$, respectively. Since MV algorithm is used in the proposed method, twice, the effects of length of $L$ has been investigated, and the results showed that it effects MVB-DMAS the same as it effects MV. The proposed algorithm significantly outperforms DMAS and MV in the terms of resolution and level of sidelobes, respectively, mainly due to having the specifications of DMAS and MV at the same time. In fact, MVB-DMAS uses the correlation process of DMAS to suppress the artifacts and noise, and adaptive weighting of MV to improve the resolution.

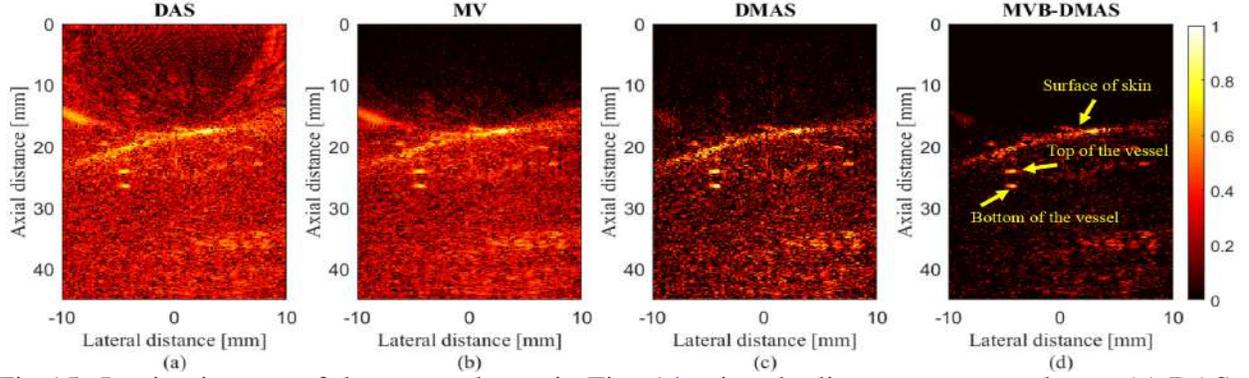

Fig 15: In vivo images of the target shown in Fig. 14 using the linear-array transducer. (a) DAS, (b) MV, (c) DMAS and (d) MVB-DMAS. All images of the median antebrachial vein. All the image are shown with a dynamic range of $60\ dB$.

## 7 Conclusion

In PAI, DAS beamformer is a common beamforming algorithm, capable of real-time imaging due to its simple implementation. However, it suffers from poor resolution and high level of sidelobes.

To overcome these limitations, DMAS algorithm was used. Expanding DMAS formula leads to multiple terms of DAS. In this paper, we introduced a novel beamforming algorithm based on the combination of MV and DMAS algorithms, called MVB-DMAS. This algorithm was established based on the existing DAS in the expansion of DMAS algebra, and it was proposed to use MV beamforming instead of the existing DAS. Introduced algorithm was evaluated numerically and experimentally. It was shown that MVB-DMAS beamformer reduces the level of sidelobes and improves the resolution in comparison with DAS, DMAS and MV, at the expense of higher computational burden. Qualitative results showed that MVB-DMAS has the capabilities of DMAS and MV concurrently. Quantitative comparisons of the experimental results demonstrated that MVB-DMAS algorithm improves CR about 20 %, 9 % and 33 %, and enhances SNR 89 %, 15 % and 35 %, with respect to DAS, DMAS and MV.

*Acknowledgments*

This research received no specific grant from any funding agency in the public, commercial, or not-for-profit sectors, and the authors have no potential conflicts of interest to disclose.

**Moein Mozaffarzadeh** was born in Sari, Iran, in 1993. He received the B.Sc. degree in electronics engineering from Babol Noshirvani University of Technology, Mazandaran, Iran, in 2015 and M.Sc. degree in biomedical engineering from Tarbiat Modares University, Tehran, Iran, in 2017. He is currently a research assistant at research center for biomedical technologies and robotics, institute for advanced medical technologies, Tehran, Iran. His current research interests include photoacoustic image reconstruction, ultrasound beamforming and biomedical imaging.

**Ali Mahloojifar** obtained a B.Sc. degree in electronic engineering from Tehran University, Tehran, Iran, in 1988, and he received the M.Sc. degree in digital electronics from Sharif University of

Technology, Tehran, Iran, in 1991. He obtained his Ph.D. degree in biomedical instrumentation from the University of Manchester, Manchester, UK, in 1995. Dr. Mahloojifar joined the Biomedical Engineering group at Tarbiat Modares University, Tehran, Iran, in 1996. His research interests include biomedical imaging and instrumentation.

**Mahdi Orooji** was born in Tehran, Iran, in 1980. He received the B.Sc. degree in electrical engineering from the University of Tehran, in 2003; the M.Sc. degree in electrical engineering from Iran University of Science and Technology, Tehran, in 2006; and the Ph.D. degree in communication systems and signal processing from the School of Electrical Engineering and Computer Science, Louisiana State University (LSU), Baton Rouge, LA, USA, in 2013. Since June 2013, he has been with the Center for Computational Imaging and Personalized Diagnostics, Department of Biomedical Engineering, Case Western Reserve University, Cleveland, OH, USA. Dr. Orooji has served several IEEE transactions and conferences as a reviewer and joined the Biomedical Engineering group at Tarbiat Modares University, Tehran, Iran, in 2016.

**Karl Kratkiewicz** graduated summa cum laude with a B.S. in biomedical physics from the Wayne State Universitys Honors College, Detroit, Michigan, in 2017. He is now a biomedical engineering PhD student at Wayne State University. His area of research includes multimodal (ultrasound/photoacoustic/elastography) imaging and photoacoustic neonatal brain imaging.

**Saba Adabi** currently is a research fellow in medical imaging in Mayo Clinic. She received her Ph.D. in Applied Electronics department in biomedical engineering from Roma Tre university in Rome, Italy, in 2017. She joined Wayne State University as a scholar researcher in 2016 (to 2017). Her main research interests include optical coherence tomography, photoacoustic/ultrasound imag-ing. She is a member of OSA and SPIE.

**Mohammadreza Nasiriavanaki** received a Ph.D. with outstanding achievement in medical optical imaging and computing from the University of Kent, United Kingdom, in 2012. His bachelor's and master's degrees with honors are in electronics engineering. In 2014, he completed a three-year postdoctoral fellowship at Washington University in St. Louis, in the OILab. He is currently an assistant professor in the Biomedical Engineering, Dermatology and Neurology Departments of Wayne State University and Scientific member of Karmanos Cancer Institute. He is also serving as the chair of bioinstrumentation track.

# List of Figures

5    Lateral variations of MVB-DMAS at the depth of 45 $mm$ for $L$=16, $L$=32 , $L$=45 and $L$=64.

6    Images of the simulated point targets phantom using the linear-array transducer. (a) DAS+CF, (b) MV+CF, (c) DMAS+CF, (d) MVB-DMAS+CF. All images are shown with a dynamic range of 60 $dB$. Noise was added to the detected signals considering a SNR of 50 $dB$.

7    Lateral variations of DAS+CF, MV+CF, DMAS+CF and MVB-DMAS+CF at the depths of 45 $mm$.

8    Schematic of the experimental setup.

9    Photographs of the phantom used in the first experiment.

10    Experimental setup of PA linear-array imaging of two parallel wires.

11    Images of the phantom shown in Fig. 9 using the linear-array transducer. (a) DAS, (b) MV, (c) DMAS and (d) MVB-DMAS. All images are shown with a dynamic range of 60 $dB$.

12    Images of the wires shown in Fig. 10 using the linear-array transducer. (a) DAS, (b) MV, (c) DMAS and (d) MVB-DMAS. All images are shown with a dynamic range of 60 $dB$.

13    Lateral variations at the depth of 33 $mm$.

14    The photo of the median antebrachial vein of a 30 years old Middle Eastern male.

15    In vivo images of the target shown in Fig. 14 using the linear-array transducer. (a) DAS, (b) MV, (c) DMAS and (d) MVB-DMAS. All images of the median antebrachial vein. All the image are shown with a dynamic range of 60 $dB$.

# List of Tables